\documentclass[twocolumn]{aastex631}

\begin{document}

\title{X-ray polarization study of the neutron star low-mass X-ray binary GX~349$+$2}

\author{Unnati Kashyap}
\affiliation{Department of Physics and Astronomy, Texas Tech University, Lubbock, TX 79409-1051, USA} 

\author{Thomas J. Maccarone}
\affiliation{Department of Physics and Astronomy, Texas Tech University, Lubbock, TX 79409-1051, USA} 

\author{Mason Ng}
\affiliation{Department of Physics, McGill University, 3600 rue University, Montr\'{e}al, QC H3A 2T8, Canada}
\affiliation{Trottier Space Institute, McGill University, 3550 rue University, Montr\'{e}al, QC H3A 2A7, Canada}

\author{Eliot C. Pattie}
\affiliation{Department of Physics and Astronomy, Texas Tech University, Lubbock, TX 79409-1051, USA}

\author{Swati Ravi}
\affiliation{MIT Kavli Institute for Astrophysics and Space Research, Massachusetts Institute of Technology, Cambridge, MA 02139, USA}

\author{Herman L. Marshall}
\affiliation{MIT Kavli Institute for Astrophysics and Space Research, Massachusetts Institute of Technology, Cambridge, MA 02139, USA}



\begin{abstract}
We report the first X-ray polarimetric results of the neutron star (NS) low-mass X-ray binary (LMXB) Z-source GX~349$+$2 using the Imaging X-ray Polarimetry Explorer (IXPE). We discovered that the X-ray source was polarized at  ${\rm PD} = 1.1 \pm 0.3\%$ ($1\sigma$ errors) with a polarization angle of ${\rm PA} = 32 \pm6 \degr$ ($1\sigma$ errors). Simultaneous Nuclear Spectroscopic Telescope Array (NuSTAR) observations show that the source transitioned through the normal branch (NB), flaring branch (FB), and soft apex (SA) of the Z-track during our IXPE observations. The X-ray spectro-polarimetry results suggest a source geometry comprising an accretion disk component, a blackbody representing the emission from the  NS surface, and a Comptonized component. We discuss the accretion geometry of the Z source in light of the spectro-polarimetric results.

\end{abstract}



\section{Introduction} \label{sec:intro}
A small population of neutron star (NS) low mass X-ray binaries (LMXBs) is persistently accreting, exhibiting distinct states defined by the spectro-temporal properties \citep{2001AdSpR..28..307B,2004astro.ph.10551V,2007A&ARv..15....1D,2007ApJ...667.1073L}. Based on spectral and temporal behavior, these NS-LMXBs are classified as Z and atoll sources \citep{1989A&A...225...79H}. The Z-sources showing Z-shaped tracks in the Hardness Intensity Diagrams (HIDs) consist of three branches, called horizontal branch (HB), normal branch (NB), and flaring branch (FB). The six known persistent Z sources in the Milky way are further classified into two sub-classes: Cyg-like (Cyg X-2, GX 5–1, and GX 340+0) and Sco-like (Sco X-1, GX 17+2, and GX 349+2) Z sources \citep{1994A&A...289..795K,1997MNRAS.287..495K,2012A&A...546A..35C,2012MmSAI..83..170C}. Additionally, detections of NS LMXBs IGR J17480-2446 \citep{2011ApJ...730L..23C,2011MNRAS.418..490C}, XTE J1701-462 \citep{2007ApJ...656..420H}, XTE J1806-246 \citep{1999ApJ...522..965W}, Cir X-1 \citep{1999ApJ...517..472S}, GX~13+1 \citep{2004A&A...418..255H}, 4U~1820$-$30 \citep{2004MNRAS.351..186M}, and 1A~1744$-$361 \citep{2024ApJ...966..232N} apparently switching between atoll and Z-like behavior have also been reported.

The Imaging X-ray Polarimetry Explorer (IXPE) has so far observed all known persistent Z-type NS LMXBs in the Milky Way: Cyg~X-2 \citep{2023MNRAS.519.3681F}, Sco~X-1 \citep{2024eas..conf..140L}, GX~5$-$1 \citep{2024AA...684A.137F}, GX~340$+$0 \citep{2024arXiv240519324B, 2024arXiv241100350B}, GX~349$+$2, and most recently GX~17$+$2 (Kashyap et al., in prep).  The previous polarization studies of Z-type NS sources indicate at least three regions that may potentially contribute to the observed polarization: the non-thermal component representing a Boundary layer (BL)/ Spreading Layer (SL) \citep{2024arXiv240916023B} or a  Comptonized plasma of hot electrons with either a shell-like or a slab-like geometry \citep{2022MNRAS.514.2561G}, the accretion disk, and the reflection of the Comptonized photons from the disk atmosphere or wind\citep{2024A&A...684A..62F,2024Galax..12...43U}. Additionally, a correlation of the polarization properties of these sources with the source position on the Z-track has also been reported for a few Z-sources (GX 5$-$1, Cyg X$-$2, and GX~340$+$0), including the transient Z-source XTE J1701$-$462 \citep{2023A&A...674L..10C}. Cir~X-1, a source exhibiting intermittent Z-source-like behavior, shows variations in the polarization angle (PA) as a function of the X-ray hardness, indicating a clear PA dependence on the spectral components \citep{2024ApJ...961L...8R}.

The Sco-like Z source GX~349$+$2 is known to be peculiar among the Z sources \citep{1998A&A...332..845K} showing only the NB or the FB branch in the HID \citep{1989A&A...225...79H, 1995NYASA.759..344K}. The NS source is known to exhibit many interesting timing features, including mHz and kHz QPOs,  broad peaked noise, and normal branch oscillation (NBO)/flaring branch oscillation (FBO) \citep{1988MNRAS.231..999P, 1998ApJ...500L.167Z, 2002MNRAS.336..217O, 2003A&A...398..223A}. Previous studies show that the continuum spectrum of GX~349$+$2  is well described by a disk blackbody component representing the disk emission, a blackbody radiation component representing the emission from the NS surface, and a non-thermal Comptonized component\citep{2018ApJ...867...64C,2023MNRAS.523.2788K}. Detections of reflection features have also been reported from GX~349$+$2 \citep{2008ApJ...674..415C, 2009A&A...505.1143I, 2018ApJ...867...64C}.

In this paper, we report the first IXPE observations of the Z type NS LMXB GX 349+2 performed from 2024 Sept 6 to 2024 Sept 9. The source was observed simultaneously with NICER and NuSTAR. In §\ref{obs}, we outline the observation details and data analysis methods. In §\ref{results}, we present the results obtained from the
spectro-polarimetric analyses. We discuss our results in §\ref{discussion} and finally summarize them in §\ref{summary}.

\section{Observations and data reductions} 
\label{obs}
\begin{figure}
\centering
\includegraphics[width=0.55\textwidth]{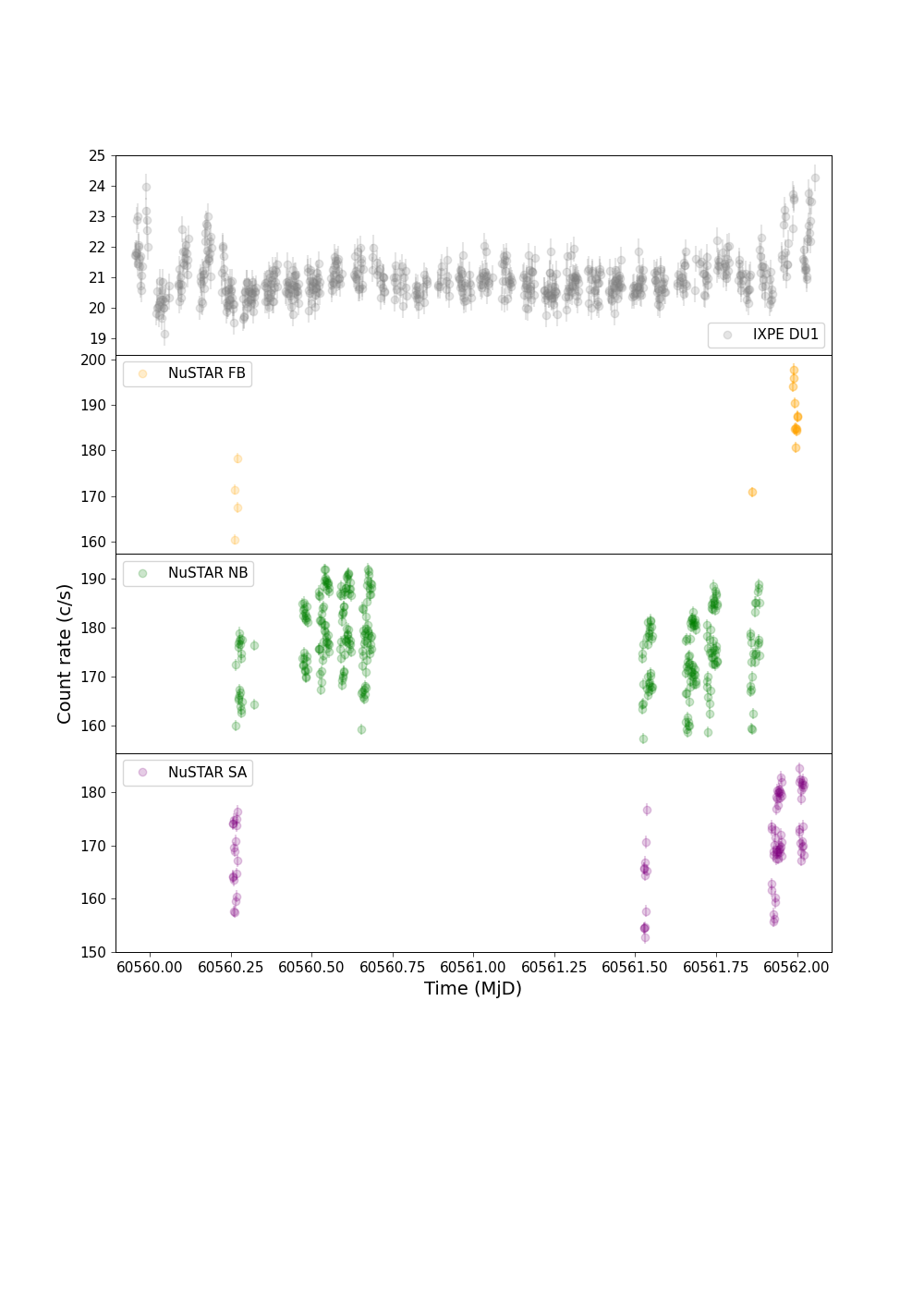}
\caption{First panel: IXPE (2-8 keV) light curve of GX~349$+$2. Second Panel: NuSTAR (3.0-79.0 keV) light curve during FB state of GX~349$+$2. Third Panel: NuSTAR (3.0-79.0 keV) light curve during NB state of GX~349$+$2. Fourth Panel: NuSTAR (3.0-79.0 keV) light curve during SA state of GX~349$+$2.
 Time bins of 128 s are used.}
\label{lc}
\end{figure}

\begin{figure}
\centering
\includegraphics[width=0.5\textwidth]{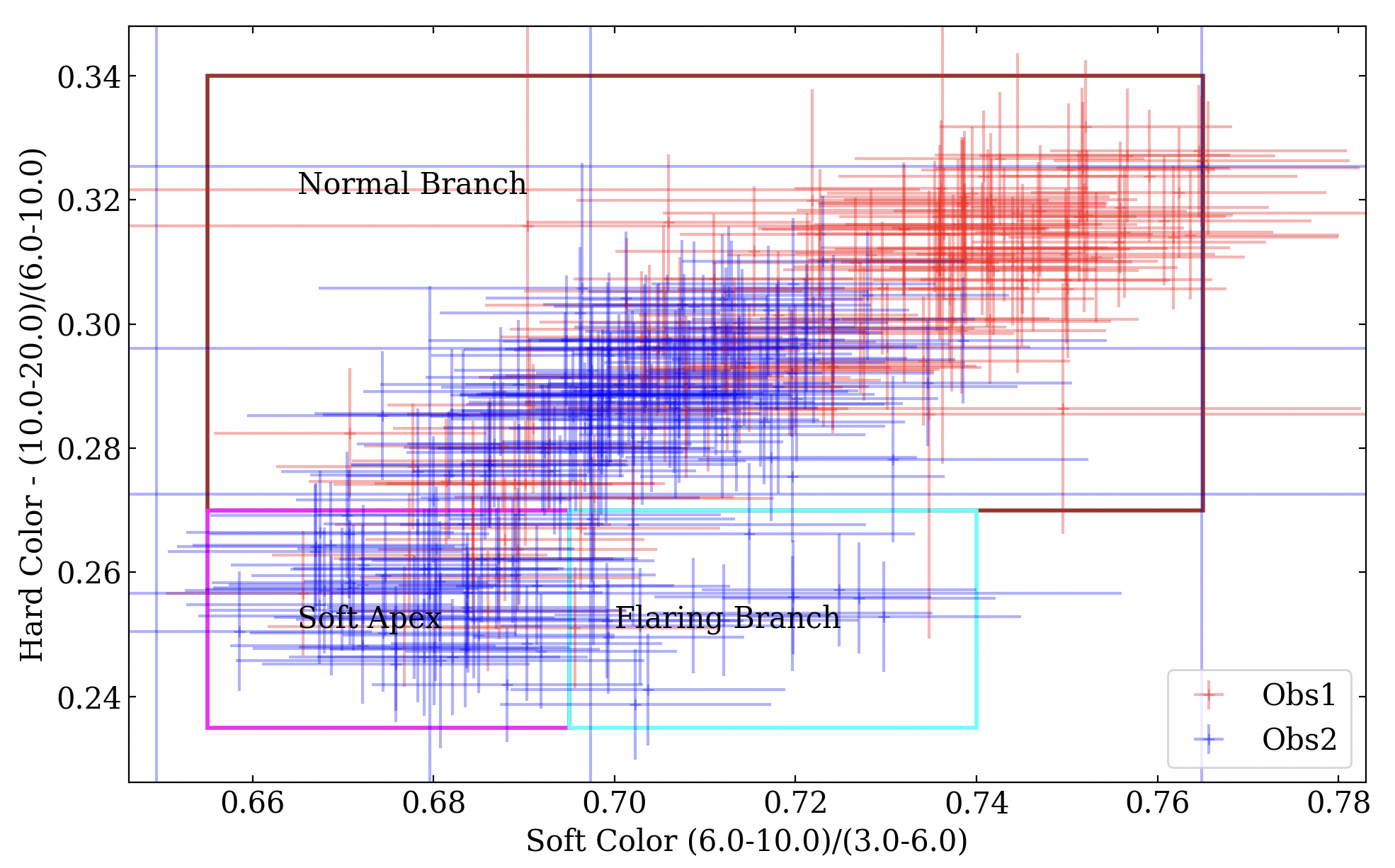}
\caption{Hardness–intensity diagram showing constructed from the two NuSTAR observations of GX~349$+$2.  Time bins of 128 s are used.}
\label{Nuhid}
\end{figure}

\begin{table*}
\centering
\caption{ IXPE, NICER, and, NuSTAR  Observations of GX 349+2 (see Section \ref{obs}). }

\begin{tabular}{c c c c c}
\hline
Instrument & Observation ID & Date (dd-mm-yyyy) & Start time (hh:mm:ss.ss) &Exposure time (ks)  \\ \hline

IXPE &03003601&06-09-2024--09-09-2024& 22:52:51.18   & 95.6 \\
NICER &7034090101-05&05-09-2024--09-09-2024 &18:43:59.00&0.3  \\
NuSTAR&9100233300[2,4] & 07-09-2024--08-09-2024& 06:05:59.43 & 17.7 \\

\hline
\label{table1}
\end{tabular}
\end{table*}

\begin{table}
\centering
\caption{Results obtained from the {\tt PCUBE} analysis. The uncertainties mentioned are 1 $\sigma$  (see Section \ref{model_ind_analysis}).}

\begin{tabular}{c c c }
\hline
Energy band& PD (\%) & PA (\degr)\\ 
\hline
2-8 keV &$1.1\pm0.3$& $32\pm 6$ \\
2-4 keV & $1.1\pm0.2$& $17\pm6$\\
4-6 keV &$0.9\pm0.4$& $49\pm12$\\
6-8 keV&$3.1\pm1.1$& $46\pm9$ \\
4-8 keV &$1.6\pm0.5$& $47\pm 8$\\
\hline
\label{table2}
\end{tabular}
\end{table}

\begin{table}
    
\caption{Results obtained from the {\tt PCUBE} analysis corresponding to the NB, SA, and FB states obtained from the simultaneous NuSTAR observations. The uncertainties are 1$\sigma$ (see Section \ref{model_ind_analysis}) }
\centering

\begin{tabular}{c c c c }
\hline
Energy band& PD (\%) & PA (\degr)&MDP$^{a}$ \\ 
\hline

FB State&&&\\
Energy band&PD (\%) & PA (\degr)&MDP\\ 
\hline

2-8 keV &$4.4\pm2.9$&$-33\pm18$ &8.8\\
2-4 keV &$3.8\pm2.7$&$-35\pm20$&8.2\\
4-6 keV &$5.2\pm4.6$&$ -23\pm 25$&14\\
6-8 keV&$7.1\pm 11.8$& $-46\pm 47$&35.9\\
4-8 keV &$5.4\pm5.1$& $-31\pm27$&15.4\\

\hline
NB State&&&\\
Energy band&PD (\%) & PA (\degr)&MDP\\ 
\hline
2-8 keV & $1.7\pm0.9$&$39\pm15$&2.8\\
2-4 keV &$1.7\pm 0.8$& $17\pm 14$&2.6\\
4-6 keV&$3.1\pm1.4$ &$52\pm 13$&4.4 \\
6-8 keV&$3.3\pm3.8$&$70\pm 32$&11.5\\
4-8 keV&$3.0\pm1.6$& $58\pm 15$&4.9 \\

\hline
SA State&&&\\
Energy band&PD (\%) & PA (\degr)&MDP\\ 
\hline
2-8 keV &  $2.4\pm1.6$& $45\pm 19$&4.9\\ 
2-4 keV &$2.3\pm 1.5$& $27\pm18$&4.5 \\ 
4-6 keV&$2.8\pm 2.6$&$69\pm 26$&7.8 \\ 
6-8 keV&$6.7\pm 6.6$&$56\pm 28$&20.0\\ 
4-8 keV &$3.9\pm2.9$& $62\pm20$&8.7 \\

\hline

\label{table3}
\end{tabular}
\footnotesize{$^a$ {\bf Minimum detectable polarization at the 99\% confidence level}}
\end{table}


\subsection{IXPE}
\begin{table*}
\centering
\caption{Best-fitting spectral model parameters from an absorbed blackbody radiation component ({\tt bbodyrad}),  multicolour disk blackbody component ({\tt diskbb}), thermally comptonized continuum component ({\tt nthcomp}) and a diskline ({\tt diskline}) model {\tt tbabs*(bbodyrad+diskbb+diskline+nthcomp)*polconst*const} to the joint NuSTAR and IXPE spectra of GX~349$+$2 during NB, SA and FB states. The uncertainties are 1$\sigma$. The calibration constant for NuSTAR FPMA is fixed at unity (see Section \ref{spec_pol_analysis}).}

\begin{tabular}{c c c c c c c c c}
\hline
Parameters& & &Spectral States & & & & \\
\hline
&NB state&&SA state && FB state&\\
\hline
&tbabs&\\
nH ($10^{22}$ atoms $cm^{-2}$) &$1.87^{+0.09}_{-0.07}$&&$2.38^{+0.18}_{-0.34}$&&$1.88^{+0.39}_{-0.40}$\\
\hline
&bbodyrad&&\\
kT (keV) &$1.29^{+0.01}_{-0.01}$&&$1.24^{+0.01}_{-0.01}$&&$1.30^{+0.02}_{-0.02}$\\
Norm&$262.12^{+4.96}_{-6.51}$&&$417.56^{+11.52}_{-19.05}$&&$381.27^{29.76}_{-31.69}$\\
\hline
&diskbb&\\
$kT_{\rm in} (keV) $&$0.74^{+0.01}_{-0.01}$&&$0.62^{+0.03}_{-0.01}$&&$0.70^{+0.06}_{-0.04}$\\
DBB Norm&$1647.70^{+174.55}_{-140.39}$&&$4626.29^{+810.96}_{-1298.05}$&&$2494.06^{+1309.77}_{-837.25}$\\
\hline
&diskline&&\\
Line E (keV) &$6.66^{+0.02}_{-0.02}$&&$6.70^{+0.02}_{-0.02}$&&$6.69^{+0.03}_{-0.03}$\\
$\beta$ &$-2.41^{+0.15}_{-0.15}$&&$-2.52^{+0.22}_{-0.32}$&&$-10.00^{+7.08}$\\
$R_{\rm in}$ ($GM/c^{2}$)&$19.40^{+5.43}_{-4.94}$&&$28.29^{+12.22}_{-8.34}$&&$80.78^{+12.02}_{-25.42}$\\
$R_{\rm out}$ ($GM/c^{2}$)&$1000.00^{a}$&&\\
Incl (\degr)&$35^{a}$&&\\
Norm ($\times 10^{-3}$) &$5.15^{+0.33}_{-0.32}$&&$7.72^{+0.90}_{-0.50}$&&$9.87^{+0.80}_{-0.81}$\\
\hline
&nthcomp&&\\
$\Gamma$&$1.33^{+0.05}_{-0.05}$&&$1.00^{+0.03}$&&$1.00^{+0.26}$\\
$kT_{\rm e}$ (keV) &$2.62^{+0.01}_{-0.01}$&&$2.42^{+0.01}_{-0.01}$&&$2.46^{+0.04}_{-0.03}$\\
$kT_{\rm BB} (keV) $&$kT_{\rm in}$&&\\
inp type&$1^{a}$&&\\
Redshift&$0^{a}$&&\\
Norm&$ 0.33^{+0.20}_{-0.12}$&&$0.01^{+0.1092}_{-0.0004}$&&$0.01^{+0.072}_{-0.001}$\\
\hline
&Cross-calibration&\\
FPMB&$0.98^{+0.0007}_{-0.0007}$&&$0.98^{+0.001}_{-0.001}$&&$0.98^{+0.002}_{-0.002}$\\
DU1&$0.84^{+0.003}_{-0.003}$&&$0.83^{+0.005}_{-0.005}$&&$0.86^{+0.009}_{-0.009}$\\
DU2&$0.84^{+0.003}_{-0.003}$&&$0.84^{+0.005}_{-0.005}$&&$0.87^{+0.009}_{-0.009}$\\
DU3&$0.82^{+0.003}_{-0.003}$&&$0.81^{+0.005}_{-0.005}$&&$0.84^{+0.009}_{-0.009}$\\
\\
$\chi^{2}$/DOF& 2747/2647&&2428/2344&&1894/2036 \\

\hline
$\text{Flux}^{b}$ ($10^{-9}$ $ergs/cm^{2}/s$)&&\\
$\text{F}_{\rm Total}$&$13.57^{+0.10}_{-0.07}$&&$13.76^{+0.37}_{-0.23}$&&$14.80^{+0.50}_{-0.43}$\\
$\text{F}_{\rm bbodyrad}$ &$6.21^{+0.16}_{-0.12}$&&$9.92^{+0.09}_{-0.12}$&&$9.10^{+2.99}_{-0.18}$\\
$\text{F}_{\rm diskbb}$ &$3.24^{+0.09}_{-0.08}$&&$3.43^{+0.16}_{-0.23}$&&$3.47^{+0.29}_{-0.28}$\\
$\text{F}_{\rm nthcomp}$ &$4.41^{+0.51}_{-0.83}$&&$2.09^{+0.15}_{-0.08}$&&$2.20^{+1.33}_{-0.21}$\\

\hline
\end{tabular}
\begin{flushleft}
\footnotesize{$^a$ fixed parameters}\\
\indent\footnotesize{$^{b}$ Energy flux of different model componenets in the 2-8 energy range } \\
\footnotesize{$\Gamma$ is found at the hard upper limit of the {\tt nthcomp} model during the SA and FB }\\
\footnotesize{$\beta$ is found at the hard upper limit of the {\tt diskline} model during the  FB }\\
\end{flushleft}

\label{table4}
\end{table*}



\begin{sidewaystable*}
\vspace{10cm} 

\caption{{ PD and PA of each spectral component obtained considering the cases where (1) PAs of the individual components are free , (2) Linked PA $\text{PA}_{\rm nthcomp}=\text{PA}_{\rm diskbb}+90 \degr$ set-up and $\text{PA}_{\rm nthcomp}=\text{PA}_{\rm diskbb}\degr$  set-up, and 3) Linked PA with a similar set-up as case (2) and (3) with an unpolarized {\tt bbodyrad} component from the best-fit spectro-polarimetric models {\tt tbabs*(bbodyrad*polconst+diskbb*polconst+diskline+
nthcomp*polconst)*const} and {\tt tbabs*(bbodyrad+diskbb+
diskline+nthcomp)*polconst*const} to the joint NuSTAR and IXPE spectra of GX~349$+$2 during NB, SA and FB states. The uncertainties reported are  1$\sigma$ (see Section \ref{spec_pol_analysis})}}

\begin{tabular}{ c c c c c c c c c c c }\\
\hline
&NB&\\
\hline
Set-up& Free PA& &Linked PA & &Linked PA&&unpolarized {\tt BB}&&unpolarized {\tt BB} \\ 
\hline
Component& PD (\%)& PA (\degr) &PD (\%)& PA (\degr) &PD (\%)& PA (\degr)&PD (\%)& PA (\degr) &PD (\%)& PA (\degr) \\ 
\hline
bbodyrad&$<$ 17.6&-&$<$14.2 &-&$2.6^{+6.7}_{-2.4}$&$<$61&0 &-&0&-\\
diskbb&$3.8^{+3.4}_{-3.4}$&$11^{+32}_{-32}$&$3.6^{+3.3}_{-3.6}$&$20^{+23}_{-23}$&$<$6.6&$17^{+22}_{-22}$&$4.0^{+2.8}_{-2.0}$&$25^{+26}_{-26}$&$2.5^{+1.7}_{-2.6}$&$34^{+27}_{-12}$\\
nthcomp&$<$ 26.2&-&$<$23.7&$\text{PA}_{\rm diskbb}-90 \degr$&$<$8.5&$\text{PA}_{\rm diskbb}$&$<$6.7&$\text{PA}_{\rm diskbb}-90 \degr$&$<$2.6&$\text{PA}_{\rm diskbb}$\\
$\chi^{2}$/DOF&2745/2643&&2745/2644&&2745/2644&&2746/2646&&2746/2646&\\

\hline
Overall&$0.7^{+0.6}_{-0.6}$&$46^{+34}_{-26}$&&&&\\
\hline
\\
&SA&\\
\hline
Component& PD (\%)& PA (\degr) &PD (\%)& PA (\degr) &PD (\%)& PA (\degr)&PD (\%)& PA (\degr) &PD (\%)& PA (\degr) \\  
\hline
bbodyrad&$11.9^{+7.7}_{-7.8}$ &$-36^{+15}_{-19}$&$<$3.1&-&$11.6^{+7.5}_{-8.0}$ &$-40^{+17}_{-19}$&0&-&0 &-\\
diskbb&$14.0^{+6.6}_{-6.7}$&$45^{+13}_{-14}$&$7.0^{+4.7}_{-4.8}$&$37^{+21}_{-17}$&$14.0^{+6.6}_{-6.6}$&$46^{+14}_{-14}$&$6.4^{+3.5}_{-2.8}$&$36^{+13}_{-13}$&$6.1^{+3.5}_{-3.6}$&$37^{+13}_{-13}$\\

nthcomp&$45.6^{+28.4}_{-28.3}$ &$53^{+18}_{-18}$&$<$50&$\text{PA}_{\rm diskbb}-90$&$44.5^{+28.8}_{-29.2}$ &$\text{PA}_{\rm diskbb}$&$<$6.8&$\text{PA}_{\rm diskbb}-90$&$<$15.3 &$\text{PA}_{\rm diskbb}$\\

$\chi^{2}$/DOF&2424/2340&&2427/2341&&2425/2341&&2427/2343&&2427/2343&\\
\hline
Overall&$2.2^{+1.1}_{-1.1}$&$37^{+15}_{-16}$&&&&\\
\hline
\\
&FB&\\
\hline
Component& PD (\%)& PA (\degr) &PD (\%)& PA (\degr) &PD (\%)& PA (\degr)&PD (\%)& PA (\degr) &PD (\%)& PA (\degr) \\ 
\hline
bbodyrad&$<$ 21.8 &-&$6.0^{+5.7}_{-5.2}$&$-52^{+52}_{-29}$&$<$ 22.1 &-&0&-&0 &-\\
diskbb&$<$ 20.2&-&$<$10.2&-&$<$ 19.6&-&$<$13.0&$-33^{+26}_{-26}$&$<11.8$ &$-27^{+24}_{-25}$\\
nthcomp&$<$85.6 &-&$<$53.6&$\text{PA}_{\rm diskbb}+90$&$<$85.8 &$\text{PA}_{\rm diskbb}$&$<$7.9&$\text{PA}_{\rm diskbb}+90$&$<$28.1 &$\text{PA}_{\rm diskbb}$\\
$\chi^{2}$/DOF&1894/2032&&1894/2033&&1894/2033&&1895/2035&&1894/2035&\\
\hline
Overall&$2.7^{+2.0}_{-2.0}$&$-29^{+22}_{-23}$&&&&\\

\hline

\end{tabular}

\label{table_comb}

\end{sidewaystable*}


IXPE observed GX~349$+$2 from 2024 Sept 6, 22:52:51.18 UTC to Sept 9, 09:57:25.184 UTC with  a total livetime of approximately 95.6 ks with a total on source exposure time of 99.2 ks for each detector unit (DU) (see Table \ref{table1} and the light curve in Figure \ref{lc}). Spectral and polarimetric analysis was performed using HEASOFT version 6.33, with the IXPE Calibration Database (CALDB) version 20240125\footnote{\url{https://heasarc.gsfc.nasa.gov/docs/ixpe/caldb/}}.  For extracting images and spectra, {\tt XSELECT} available as a part of the {\tt HEASoft 6.33} package was used extensively. Source photons were selected from a circular region with a radius of $60\arcsec$ for I, Q, and U spectra for each detector unit centered at the brightest pixel located at RA of 256$\fdg$43 and DEC of -36$\fdg$41. The weighted scheme  NEFF was adopted for the spectro-polarimetric analysis with improved data sensitivity\footnote{\url{https://heasarc.gsfc.nasa.gov/docs/ixpe/analysis/IXPE_quickstart.pdf}} \citep{2022SoftX..1901194B, 2022AJ....163..170D}.  The ancillary response files (ARFs)
and modulation response files (MRFs) were generated for each DU
using the {\tt ixpecalcarf} task, with the same extraction
radius used for the source region. GX~349$+$2 being a bright Z-source, following the prescription by \cite{2023AJ....165..143D}, we did not implement background rejection or subtraction. The unweighted model-independent polarimetric analysis was performed using the {\tt IXPEOBSSIM package version 31.0.1} \citep{2022SoftX..1901194B}.

\subsection{NICER}
The Neutron star Interior Composition Explorer (NICER) observed GX~349$+$2 from 2024 Sept 5, 18:43:59.00 UTC to Sept 9, 00:24:00.00	UTC. The observation details are summarized in Table \ref{table1}.  We note that a major fraction of the NICER observations were carried out during orbit day observations severely affected by the optical light leak, and the filtered data comprised only 375 s on-source exposure time. Hence, we did not use NICER observations for the spectro-polarimetric analysis.

\subsection{NuSTAR}

The Nuclear Spectroscopic Telescope Array (NuSTAR) is a hard X-ray focusing telescope operating over 3--79~keV. NuSTAR observed GX~349+2 over two Target of Opportunity observations (ObsIDs 91002333002 and 91002333004) across 2024 September 7 and 8, for a total of 17.7~ks of filtered exposure (see Table~\ref{table1}). The data were reduced with the \texttt{nupipeline} tool from the NuSTAR Data Analysis Software (NuSTARDAS) v2.1.4 from HEASoft v6.34 with the calibration database dated 20240325. The source photons were extracted from a circular region centered at (R.A., Dec.) = ($256\fdg4357$, $-36\fdg4230$) with a radius of $120\arcsec$, while an estimate of the background was made from a source-free region with radius $120\arcsec$ near the source, centered at (R.A., Dec.) = ($256\fdg3547$, $-36\fdg5576$). The scientific products, such as energy-dependent light curves and spectra (including branch-resolved products; see Figure~\ref{Nuhid}), were generated with \texttt{nuproducts}. The light curves were binned with 128 s time bins, and the spectra were regrouped such that there were 30 counts per spectral bin.
%


\section{Results}
\label{results}
\subsection{Source Spectral State}
Figure~\ref{lc} shows the IXPE and NuSTAR light curves of GX~349$+$2 during our observations. The patterns of the Z-track of the source were not discernible in the IXPE hardness intensity diagram (HID) or color-color diagram (CCD), due to its limited energy coverage, hence, we used simultaneous NuSTAR observations to track the spectral state transitions of GX~349$+$2 during our observations. Figure~\ref{Nuhid} shows the NuSTAR color-color diagram (CCD), where the hard and soft colors are defined as the ratios of the count rates in the energy bands between 10.0-20.0 keV and 6.0-10.0 keV, and between 6.0-10.0 keV and 3.0-6.0 keV, respectively. The NuSTAR CCD indicates the source transition through NB, SA (soft apex), and FB throughout the NuSTAR pointings (obs1 and obs 2). To further investigate the source behavior during our observations, we performed detailed spectro-polarimetric investigations for each of the states separately.

\subsection{Model-independent polarimetric analysis}
\label{model_ind_analysis}

\begin{figure}
\centering
\includegraphics[width=0.55\textwidth]{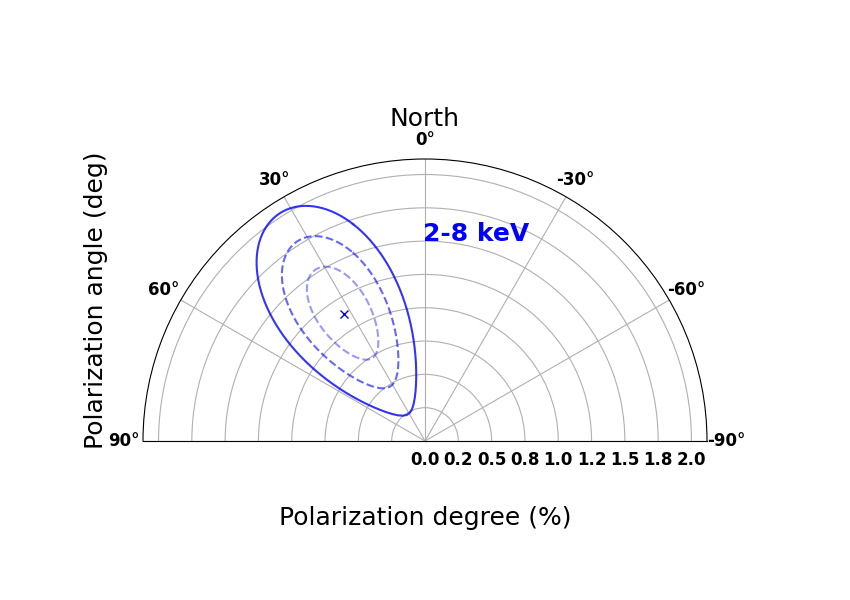}
\includegraphics[width=0.55\textwidth]{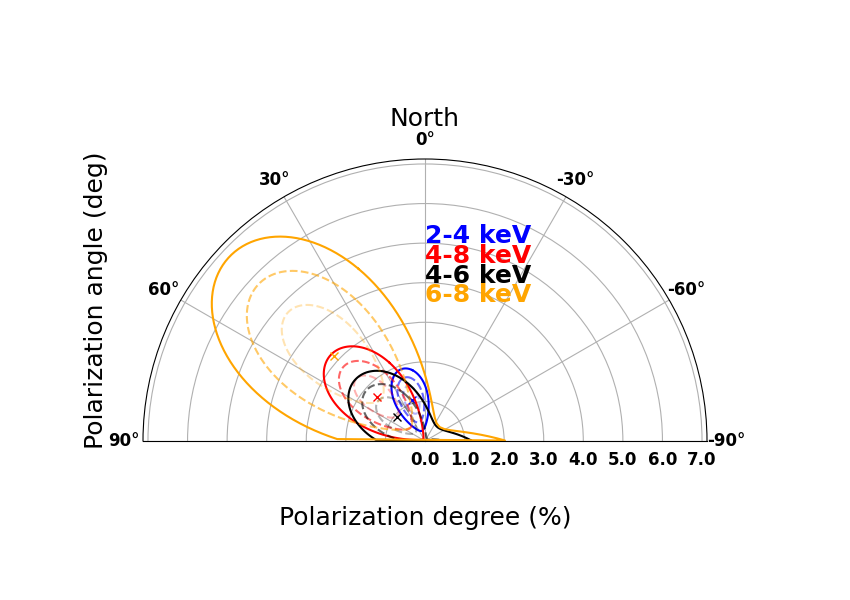}

\caption{Left panel: Contour plots of the polarization degree and angle, determined with the {\tt PCUBE} algorithm, at the 68 \%, 95 \% and 99.7 \% confidence levels, in the 2–8 (upper panel, blue) energy band. Right panel: Contour plots of the polarization degree and angle, determined with the {\tt PCUBE} algorithm, at the 68 \%, 95 \% and 99.7 \% confidence levels, in the 2–4 keV (blue), 4-8 keV (red), 4-6 keV (black), and 6-8 keV (orange) energy bands.  }
\label{mod_ind_pol}
\end{figure}

\begin{figure}
\centering
\includegraphics[width=0.37\textwidth]{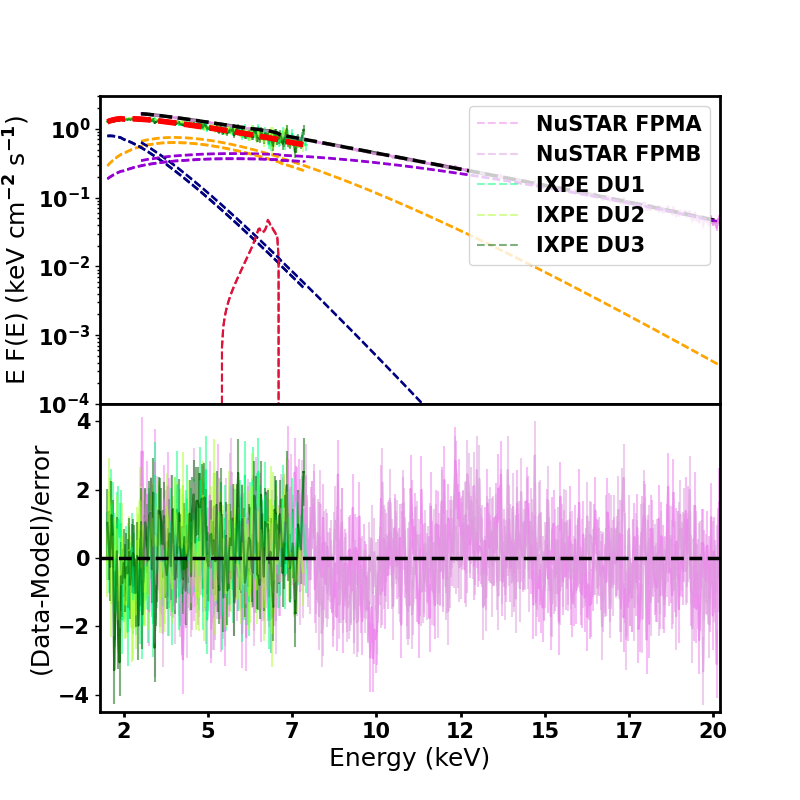}
\includegraphics[width=0.37\textwidth]{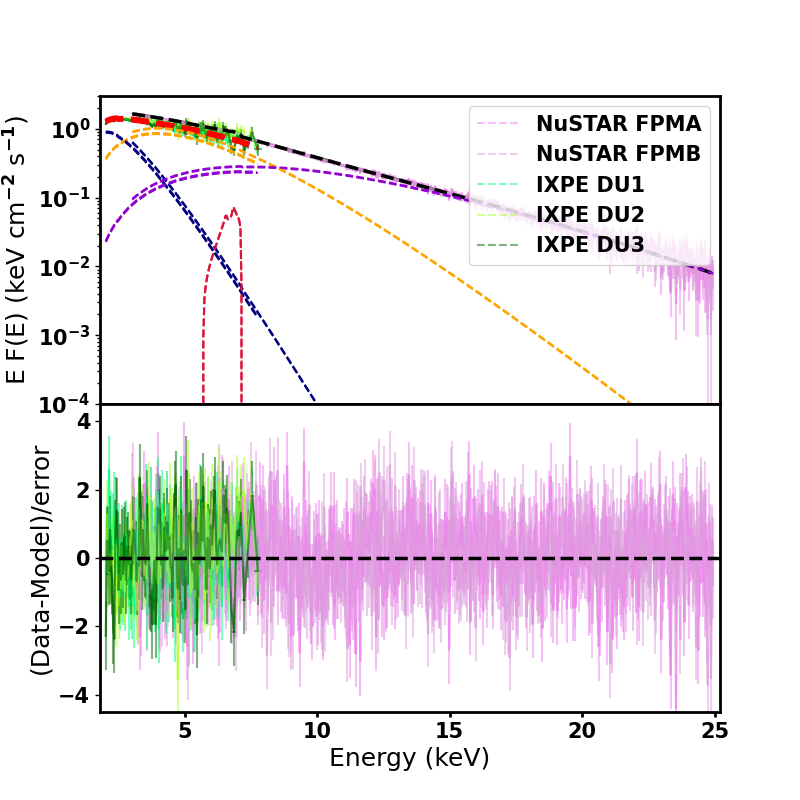}
\includegraphics[width=0.37\textwidth]{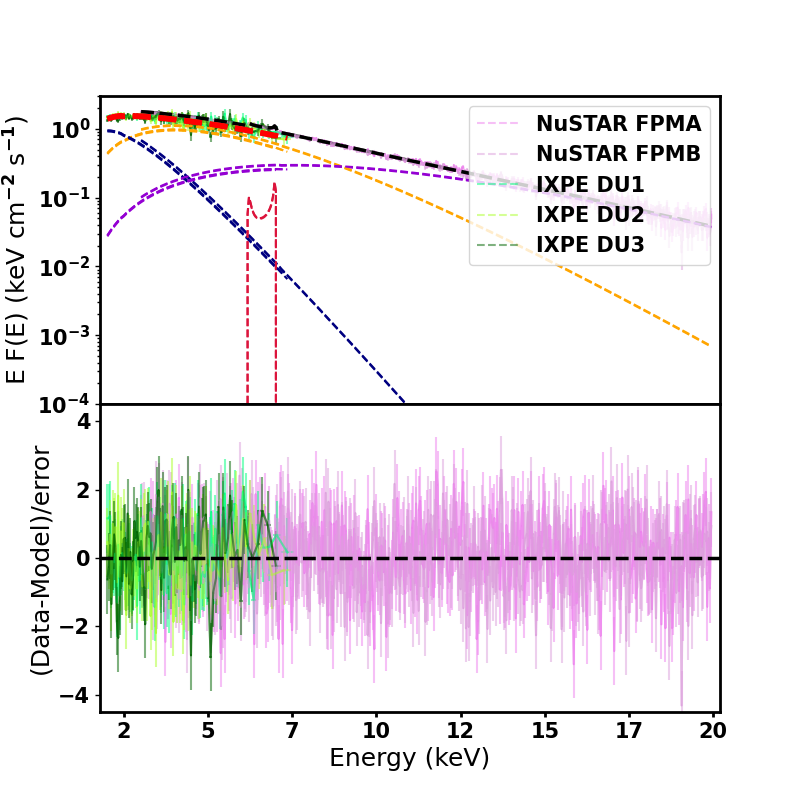}

\caption{Model fitted deconvolved joint spectra of GX~349$+$2 as observed by IXPE DU1 (spring green), IXPE DU2 (green yellow), IXPE DU3 (dark green), and NuSTAR (magenta). The spectra are fitted with the {\tt tbabs*(bbodyrad+diskbb+diskline+nthcomp)*polconst
*const} model in the 2-30 keV (NB, top), 2-25 keV (SA, middle), and 2-20 keV (FB, bottom) energy band. The total model is shown with the dashed black (NuSTAR) and red line (IXPE), and the individual additive components {\tt bbodyrad}, {\tt diskbb}, {\tt diskline}, and {\tt nthcomp} are shown with the dashed orange, dashed navy, dashed crimson, and dashed violet lines, respectively. The lower subpanel shows the residuals between the data and the best fit model.}
\label{spec_all}
\end{figure}

\begin{figure}
\centering
\includegraphics[width=0.45\textwidth]{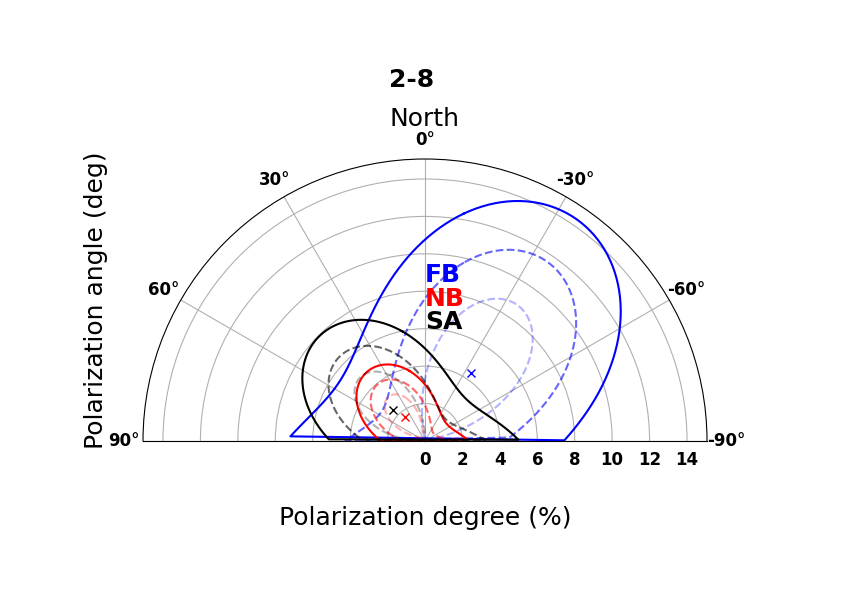}
\includegraphics[width=0.45\textwidth]{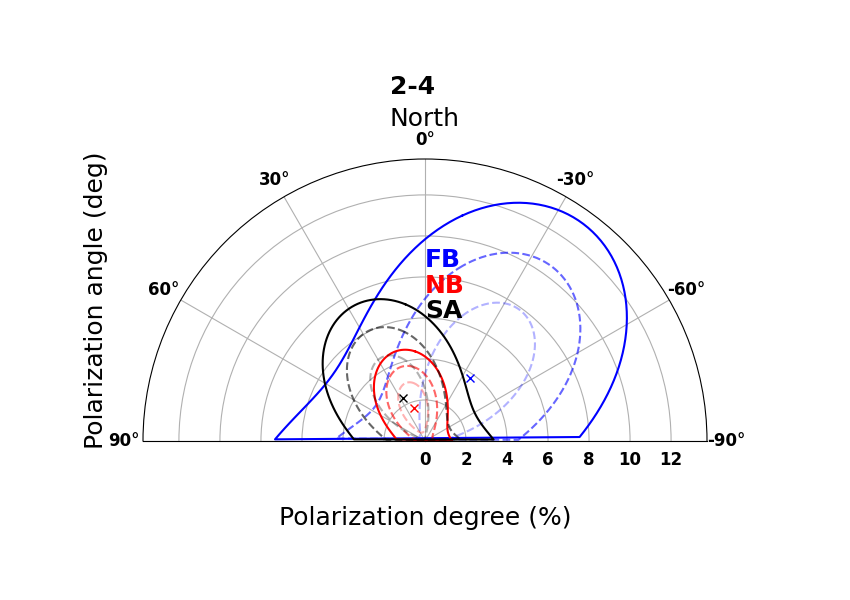}
\includegraphics[width=0.45\textwidth]{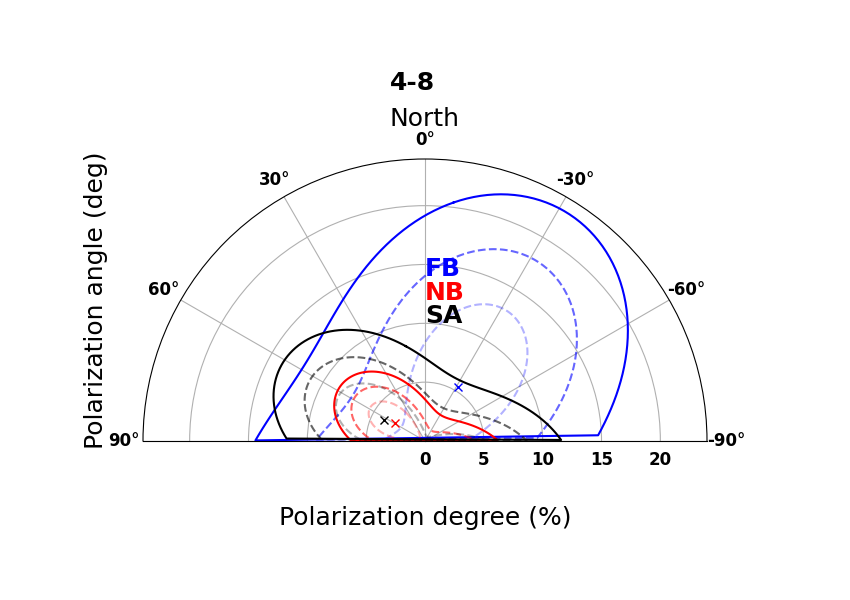}
\includegraphics[width=0.45\textwidth]{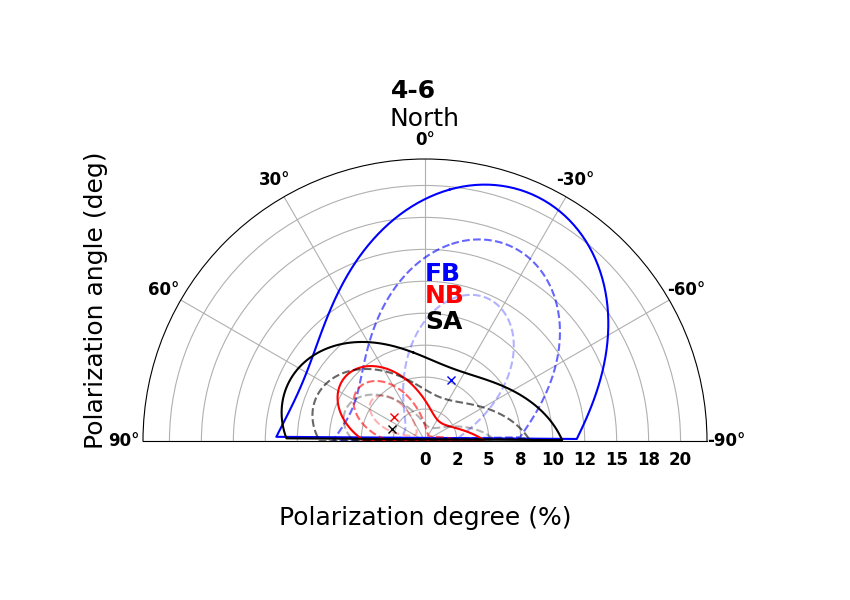}

\caption{Contour plots of the polarization degree and angle in the FB (blue), NB (red), and SA (black), determined with the {\tt PCUBE} algorithm, at the 68 \%, 95 \% and 99.7 \% confidence levels, in the 2–8 (first panel), 2–4 (second panel), 4–8 (third panel), 4–6 (fourth panel) energy band in the FB (blue), NB (red), and SA (black) state.}
\label{mod_ind_pol_br}
\end{figure}

We employed polarimetric analysis of GX~349$+$2 using the {\tt ixpeobssim} package \citep{2022SoftX..1901194B} under the {\tt PCUBE} algorithm in the {\tt xpbin} tool. We applied unweighted analysis implemented in the {\tt ixpeobssim} package and employed the polarimetric analysis in the 2–8 keV, and 2-4 keV, 4-6 keV, 6-8 keV, and 4-8 keV energy bands. The results obtained are reported in Table \ref{table2} and Figure \ref{mod_ind_pol}, which shows a significant detection of polarization ${\rm PD} = 1.1 \pm 0.3\%$ ( significance at 3.9$\sigma$) with a polarization angle of ${\rm PA} = 32 \pm6 \degr$ from GX~349$+$2 in the 2-8 keV energy band. We also detected a higher polarization (considering 1$\sigma$ uncertainty) of ${\rm PD} = 3.1 \pm 1.1\%$ ( significance at 2.5$\sigma$) with a polarization angle of ${\rm PA} = 46 \pm9 \degr$ in the 6-8 keV energy band. \\

 As it is clear from the NuSTAR CCD that the source shows transitions through different Z-branches during our IXPE observations, to estimate source polarization during each of the spectral states separately, we extracted IXPE data for the time intervals that overlap with the intervals of each of the Z-branches obtained from the NuSTAR (obs 1 and obs 2) CCD. First, we extracted the NuSTAR GTIs corresponding to the NB, SA, and FB branches for obs 1 and obs 2. We then filtered IXPE data selecting only the intervals that overlap with NuSTAR observations (obs1 and obs2) corresponding to each of the branches using the {\tt XSELECT} package. The total IXPE exposures that fall within the NuSTAR NB, SA, and FB states are 34.7 ks, 596.0 s, and 0 ks during the obs 1 and 34.3 ks, 43.7 ks, and 30.5 ks during the obs 2, respectively.

Finally, we performed a model-independent polarimetric analysis of each of the individual Z-branches. The results obtained from the branch-resolved model-independent polarimetric studies are reported in Table~\ref{table3}, which indicates no significant polarization detection in any of the branches or energy bins analyzed.
 The contour plots of the branch-resolved PD and PA in the 2-8 keV, 2-4 keV, 4-8 keV, and 4-6 keV are represented in Figure \ref{mod_ind_pol_br}, which hints towards a PA rotation by $\sim$ 60\degr  between the SA/NB and FB state of the source in all the energy bands, albeit with large uncertainties. Further IXPE observations with simultaneous NuSTAR observations with longer exposure times are necessary for investigating the polarization of the source, uniquely identifying properties associated with each of the branches.

\subsection{Spectro-polarimetric analysis}
\label{spec_pol_analysis}

 To carry out the spectro-polarimetric analysis, we first extracted the IXPE I, Q, and U spectra from the same IXPE (NB, SA, and FB) observations filtered during the model-independent analysis (see section \ref{model_ind_analysis}). We then performed the spectral fitting and statistical analysis, using the {\tt XSPEC v 12.14.0h} spectral fitting package distributed as a part of the {\tt HEASoft 6.33 package}. Considering the IXPE and the NuSTAR FPMA and FPMB observations, we fitted the spectra of each of the states, namely, NB, SA, and FB with a model consisting of multicolor disk blackbody component ({\tt diskbb} in {\tt XSPEC}; \cite{1984PASJ...36..741M}), blackbody radiation component ({\tt bbodyrad} in {\tt XSPEC}), thermally Comptonized continuum component ({\tt nthcomp} in {\tt XSPEC}; \cite{1996MNRAS.283..193Z,1999MNRAS.309..561Z}), and a diskline ({\tt diskline} in {\tt XSPEC}; \cite{1989MNRAS.238..729F}) component representing the iron (emission) feature detected at $\sim$ 6.7 keV. We performed the spectral fitting in the 2.0–8.0 keV energy range (IXPE) and 3.0-20.0 keV (NuSTAR FB), 3.0-25.0 (NuSTAR SA), and 3.0-30.0 keV (NuSTAR NB) energy range to reduce the effect of background systematics in the NuSTAR spectrum. For the fitting process, we fix the outer disc radii at 1000 $R_{g}$ and inclination at 35\degr \citep{2018ApJ...867...64C}, as otherwise, the parameters become completely unconstrained. The spectrum was modified by the presence of neutral hydrogen absorption in the interstellar medium, and this was modeled with the {\tt tbabs} model. The abundances and photoelectric cross-sections are adopted from \citep{2000ApJ...542..914W}. A constant ({\tt const}) model was used to account for the uncertainties in cross-calibration uncertainties between NuSTAR FPMA, FPMB, and the IXPE DUs and is reported in Table~\ref{table4}. 

To study the polarization of the spectral components during the NB, FB, and SA, we applied the {\tt polconst} model to the entire continuum model to check the consistency with the results obtained from the model-independent analysis. The PD and the PA obtained from the spectro-polarimetric analysis are consistent with the PD and PA obtained from the model-independent analysis (see Section \ref{model_ind_analysis} and Table~\ref{table3} and \ref{table_comb}). Figure~\ref{spec_all} shows
the NuSTAR (magenta) and IXPE (spring green, green
yellow, and dark green) spectra along with the best-
fitting models, and the corresponding best fitting values are reported in Table \ref{table4} and \ref{table_comb}. Finally, we applied a model that assumes different constant polarization for each spectral component. However, the PD of the components remains unconstrained and only the upper limits on the polarization estimations of the individual components could be obtained due to limited data sensitivity. We report the PD of the individual components in Table \ref{table_comb}. 

To reduce the statistical uncertainties further and have better constraints on the PA, we consider two plausible scenarios: 1) the disk polarization being parallel and 2) the disk polarization being perpendicular to the polarization attributed to the comptonization region; i.e., we allowed to vary only the PA of the {\tt diskbb} component with the PA of {\tt nthcomp} being linked by a relation $\text{PA}_{\rm nthcomp}$=$\text{PA}_{\rm nthcomp}$ and $\text{PA}_{\rm diskbb}$=$\text{PA}_{\rm nthcomp}$+/-$90 \degr$.  However, we could only obtain an upper limit on the PD of the individual components, in most of these cases. Then we consider a third scenario, where there is no polarization from the blackbody emission ($\text{PD}_{\rm bbodyrad} = 0$), considering the fact that the blackbody is possibly attributed to  a symmetrical emission from the surface of the NS. Although considering an unpolarized blackbody emission we could only obtain upper limits for the individual components, however, the upper limits obtained on the polarization estimates of the individual components decreases slightly (see Table \ref{table_comb}). Since statistically, we could not differetiate one scenario over the other, we report the results obtained from the spectro-polarimetric analysis considering all the aforementioned cases in  \ref{table_comb}. The overall PD and the PA obtained from the branch-resolved spectro-polarimetry, are consistent with the PD and PA obtained from the model-independent analysis (see Section \ref{model_ind_analysis} and Table~\ref{table_comb}) with an indication of the PA rotation of $\sim$ $60\degr$ between the FB state (considered with 1$\sigma$ uncertainties, albeit to be taken cautiously) and the NB/SA state.

We also tried fits with more complex models, including {\tt relxillNS} and {\tt relconv*reflionx} models, to describe the reflection component. However, given the data quality, we could not obtain well-constrained parameters. Hence, we use a {\tt diskline} model to describe the iron (emission) feature detected at $\sim$ 6.7 keV.

\section{Discussion}
\label{discussion}

\begin{table*}
\centering
\scriptsize
\caption{The X-ray polarization properties of the Z-type NS LMXBs and transient Z sources observed by IXPE during different spectral states, including polarization degree (PD), polarization angle (PA), and the X-ray PA with respect to the radio jet position angle detected from the sources (see Section \ref{discussion}). }
\begin{tabular}{ c c c c c c c}
\hline
Source (Type) & PD  (\%)$^{a}$ & State & PA (\degr) &PD/PA variation &X-ray PA w.r.t.  & ref \\ 
&(2--8 keV)& & & with energy&Radio jet&\\
\hline
&& & Sco-like Z sources & &&\\

Sco~X-1  &$1.0\pm0.2$ {\bf (2$\sigma$)} &SA/FB& $8\pm6$& No&46\degr&\cite{2024ApJ...960L..11L}\\
GX~349$+$2  &$1.1\pm0.3$ {\bf (1$\sigma$)} &NB,SA,FB& $32\pm6$& No&-&This work\\
\hline
&& & Cyg-like Z sources & &&\\
GX~5$-$1 &$3.7\pm0.4$ {\bf (2$\sigma$)}& HB &$-9\pm3$ &PA&Aligned$^{b}$&\cite{2024AA...684A.137F}\\
 &$1.8\pm0.4$ {\bf (2$\sigma$)} &NB/FB&$-9\pm6$ &PA&&\cite{2024AA...684A.137F}\\
 GX~340$+$0 &$4.02\pm 0.35$ {\bf (1$\sigma$)}& HB &$37.6 \pm 2.5$&PA &&\cite{2024arXiv240519324B}\\
& $1.22\pm0.25$ {\bf (1$\sigma$)}& NB & $38\pm6$ & PD  &  &\cite{2024arXiv241100350B} \\
Cyg~X-2  &$1.8\pm0.8$ {\bf (3$\sigma$)}& NB&$140\pm12$ &PD &Aligned&\cite{2023MNRAS.519.3681F}\\
\hline
\label{table8}
\end{tabular}
\begin{flushleft}
\footnotesize{$^a$ Confidence intervals of uncertainties are quoted in parenthesis.}\\ 
\footnotesize{$^b$ Pattie et al. in prep; from recent Very Large Array radio data, the preliminary intrinsic radio position angle of GX~5--1 was found to be $\sim -18^{\circ}$, aligned in particular with the higher energy X-ray polarization angle.} 

\end{flushleft}

\end{table*}

In this paper, we report the first X-ray polarization study of the NS Z source GX 349+2 using IXPE. The simultaneous NuSTAR observation shows that the source transitions through NB, SA, and FB of its Z-track during our IXPE observations. The X-ray spectra of GX~349$+$2 is well described by three component model \citep{2007ApJ...667.1073L}, consistent with the previously reported studies of this source \citep{2001ASPC..251..400R,2004ESASP.552..317F,2018ApJ...867...64C,2023MNRAS.523.2788K}, NS sources \citep{2002MNRAS.337.1373G,2007A&ARv..15....1D} and the recent studies of Z sources \citep{2024arXiv241100350B}. The spectral studies show that the softer component is represented by the accretion disk emission with temperatures $\sim$ 0.62-0.74~keV as the source transitions through different states, while a blackbody radiation component with temperature $\sim$ 1.29–1.30 keV represents the emission possibly coming from the NS surface, with an estimated radius of
$\sim$ 14.9-18.8 km (considering a source distance of 9.2 kpc \citep{2002A&A...391..923G}) of the emitting region. The third component is described by a Comptonization of the disk seed photons in a hot plasma with electron temperature $\text{kT}_{e} \sim$ 2.42-2.62 keV. We also detect an emission feature at $\sim$ 6.7 keV, which is also reported in the previous spectral studies of the source \citep{2009A&A...505.1143I,2018ApJ...867...64C}. We note here that the discrepancy observed between the inner disk radius estimates obtained from the {\tt diskbb} ({\tt diskbb} normalization) and {\tt diskline} model fits (also reported in \cite{2023MNRAS.523.2788K}) can be explained by the fact that the {\tt diskbb} gives an apparent inner disc radius,  \citep{1998PASJ...50..667K} with correction factors that rely on disk atmosphere modeling that has, to date, focused on accreting black holes  \citep{2006ApJ...647..525D} and the {\tt diskline} model takes relativistic effects into consideration. Additionally, disk reflections may depend on many other factors, including disk ionizations, thermodynamics, geometry, or inclination angles \citep{1989MNRAS.238..729F}. We note here that the limited sensitivity of the data presented in this work limits obtaining detailed quantitative conclusions on the source geometry from the branch-resolved spectro-polarimetric parameters of the individual model components, considering each of the Z branches. Hence, we focus only on the overall understanding of the source geometry considering each of the branches. 

The model-independent polarimetric analysis of GX~349$+$2 shows a total (2-8 keV) polarization of ${\rm PD} = 1.1 \pm 0.3\%$  (significance at 3.9$\sigma$) with a ${\rm PA} = 32 \pm6 \degr$ in the 2-8 keV energy band. We also observe a hint of higher polarization ${\rm PD} = 3.1 \pm 1.1\%$ with a polarization angle of ${\rm PA} = 46 \pm 9 \degr$ in the 6-8 keV energy band (considering 1$\sigma$ uncertainties only). Additionally, the spectral studies reveal the presence of reflection features at $\sim$ 6.7 keV in the NuSTAR spectrum of the NS source (see Table \ref{table4}). Thus, the high PD (${\rm PD} = 3.1 \pm 1.1\%$, significance at 2.5$\sigma$) detected in the 6-8 keV energy band may indicate the contribution of reflected emission to the observed total PD from the source.

 \subsection{A comparison between the Cyg-like and Sco-like Z sources}
 
 The previous IXPE studies report that the polarizations in the case of NS Z sources are associated with the accretion disk, Comptonization of the seed photons in a BL/SL \citep{2024arXiv240916023B}  or a slab/shell-like corona \citep{2022MNRAS.514.2561G}), and/or the reflection or scattering in the accretion disk or the wind. We report the polarization estimation of both Cyg-like and Sco-like Z sources reported from the previous IXPE studies in Table \ref{table8}.  Sco~X-1, observed in the SA/FB state, shows a total polarization of $1.0 \pm 0.2\%$ \citep{2024ApJ...960L..11L}, with no evidence of any significant variation of polarization with energy. Moreover, the PD associated with the Comptonization region in the case of Sco~X-1 is estimated as ${\rm PD}=1.3\pm0.4\%$, shows a PA rotation by 46\degr with respect to the known radio jet position angle of the source \citep{Fomalont2001ApJ...558..283F}. This estimation is not consistent with the previous studies indicating alignment of the X-ray PA with the radio jet from Sco~X-1 \citep{1979ApJ...232L.107L,2022ApJ...924L..13L}. The rotation of PA reported by the IXPE studies of Sco~X-1 is proposed to be associated with the evolution of the geometry of the Comptonization region over time \citep{2024ApJ...960L..11L}. 

\cite{2024ApJ...960L..11L} also propose that a  changing jet orientation due to the Lense–Thirring precession of the accretion disk may be a possibility for the observed misalignment between the X-ray PA and the radio jet position angle.  We find this possibility to be rather unlikely for several reasons. First, the Lense-Thirring precession takes place on a sub-second timescale which is much quicker than the physically large scale of the jet can respond to, such that the axis of the jet will be only minorly affected, if at all, by this effect regardless of the maximum Lense-Thirring precession angle \citep{Ingram2015}. IXPE similarly observes the average polarized emission during Lense-Thirring precession episodes due to the timescales involved.

Second, the expected changes in the radio jet position angle associated with the Lense–Thirring precession of the accretion disk in the case of NS LMXBs are not consistent with the large jet-PA misalignment of 46\degr observed in the case of Sco~X-1. A 4-year resolved radio image monitoring campaign of Sco~X-1 shows that the long-term variation in the jet axis orientation of Sco X-1 is less than about 3\degr \citep{Fomalont2001ApJ...558..283F}, and the jet is unresolved in the direction perpendicular to its motion; large scale precession would lead to a large jet opening angle. Additionally, a modeling of X-ray QPOs from binary systems suggests that the NS/accretion disk interaction could produce only a 5\degr precession over several days \citep{2000ApJ...542L.111T}. The inconsistency of such estimates and observational data with the very high value of PA misalignment (46\degr) reported in the case of Sco~X-1 implies that the Lense-Thirring precession is most likely not associated with the observed PA misalignment as reported in \cite{2024ApJ...960L..11L}.

Two alternative possibilities exist. Firstly, this may be due to the evolution of X-ray polarimetric properties, potentially correlated with Z source branches, and the requirement to average over long amounts of time to make a detection means that this cannot be readily seen.  Alternatively, the different spectral components may have different polarization position angles, all either parallel or perpendicular to the jet, so that their sum is at some intermediate position angle. However, it is challenging to test this hypothesis solely with IXPE studies because the standard spectral components have similar spectral shapes over the IXPE bandpass.

As a follow-up to this work, we have planned radio observations of GX~349$+$2 with the Very Large Array in the near future to obtain the linear polarization properties of the jet and determine its intrinsic position angle.  This will help us investigate if such misalignments of X-ray and jet position angles are inherent to Sco-like sources or are peculiar to Sco X-1.

On the other hand, in the case of Cyg~X-2, observed during the NB state, the Comptonized emission with a ${\rm PD}=4.0\pm0.7\%$ is reported to be aligned with the radio jet, suggesting a disk PA orthogonal to the PA of the Comptonization region \citep{2023MNRAS.519.3681F}, and this appears to also be true for the Cyg-like source GX~5$-$1 based on the preliminary radio results (Pattie et al. in prep.), and the detailed analysis will be reported in our future work. The energy-dependent polarization study of Cyg X-2 shows a hint of marginal increase of the PD with energy which is interpreted as the presence of a softer accretion disk with weaker polarization (and/or at a right angle) than the harder BL/SL component representing the Comptonized emission. An indication of marginally lower polarization in the softer energy band than the harder energy band has also been reported from the NB state observation of Cyg-like Z source GX~340$+$0 \citep{2024arXiv241100350B}. Furthermore, the HB state energy-resolved polarization study of GX~340$+$0 shows that the lowest energy bin (i.e. 2–2.5 keV) has a different PA as compared to the PA of the rest of the emission, indicating the contribution from a third softer component (i.e. accretion disk) resulting in the observed marginal changes in the PA \citep{2024arXiv240519324B}. A deviation of the disk PA  by 59-75\degr with respect to the PA attributed to the Comptonized emission has also been reported {\bf \citep{2023A&A...674L..10C,2024AA...684A.137F}}, indicating it as an intrinsic property of the HB state Cyg-like Z-sources. However, such properties have not been observed or reported for any Sco-like sources either due to the lack of HB state observations or sensitive observations.

GX~349$+$2, being a Sco-like Z source, shows a total measured polarization (${\rm PD} = 1.1 \pm 0.3\%$) in the 2-8 keV energy band, consistent with the estimated polarization in the case of Sco~X-1 (see Table \ref{table8}). As reported in Table \ref{table2} and Figure \ref{mod_ind_pol}, the energy-dependent study, in the case of GX~349$+$2, does not show any strong variations in the PD, which is also reported in the case of Sco~X-1. Thus,  the Sco-like sources (Sco~X-1 and also GX~349$+$2) not exhibiting any such PD variations with energy may indicate an inherent difference between the Sco-like and Cyg-like Z sources possibly associated with the accretion geometry or mechanism. Our future studies of GX~17$+$2 will provide further scope to investigate the properties of the Sco-like sources collectively. 

The major difference between the Sco-like and Cyg-like sources is the presence of the stronger FB in the Sco-like sources. The flaring branch in the Cyg-like sources is known to be associated with energy release at the NS surface due to unstable nuclear burning \citep{2012A&A...546A..35C}. On the contrary, the flaring in the case of the Sco-like sources is associated with the combination of unstable burning and an increase in mass accretion rate. Moreover, unlike the Cyg-like sources, the HB is often absent in the Sco-like sources \citep{2014MNRAS.438.2784C}. However, what drives the difference in the polarization, and how that correlates with the aforementioned factors is still not understood fully due to the lack of sampling and as each class includes only a handful of sources. Thus, future investigations may help address the observed polarization difference between the Sco-like and Cyg-like sources, identifying them uniquely, including more sensitive observations with longer IXPE exposures, along with simultaneous observations with the other X-ray telescopes with broad energy coverage allowing a more accurate source state identification.

 As stated before, there is a possibility that the PA estimated with IXPE to be the total summed PA of different X-ray emission components, which can be either parallel or perpendicular to the radio jet. It is difficult to test this hypothesis solely using IXPE data, but combining results with those from future polarization missions like XPoSAT\footnote{\url{https://www.isro.gov.in/XPoSat_X-Ray_Polarimetry_Mission.html}}, or proposed missions like REDSOX\footnote{\url{https://ait.mit.edu/instruments/redsox-polarimeter/}} or PolSTAR\citep{2016APh....75....8K}, which cover both hard and soft energy ranges, where the spectral components diverge more, could test this possibility. Furthermore,  such a joint study will provide scope for broadband spectro-polarimetry shedding light on the geometry of the NS sources disentangling the softer and harder emission components enabling more precise energy-dependent polarization studies.

\subsection{Polarization variations along the Z-track}
 
Studies report variations of the polarimetric properties along the Z-track with a relatively higher PD in the HB state sources like XTE~J1701$-$462 \citep{2023A&A...674L..10C} and GX~5$-$1 (see Table \ref{table8}). XTE~J1701$-$462  showed the presence of PD variations from $4.6 \pm 0.4\%$ in the HB to $\sim$ 0.6\% in the NB. GX~5$-$1 shows a PA variation between the disk and Comptonization region, with a higher PD in the HB relative to the NB/FB . A similar variation of PD has also been reported from the polarization study of GX 340+0 (see Table \ref{table8}). Sco~X-1, observed in the SA/FB, shows a relatively lower polarization of $1.0 \pm 0.2\%$, further emphasizing the PD variation along the Z track.

As reported in Table \ref{table_comb} the branch-resolved spectro-polarimetric studies of GX~349$+$2 shows a  ${\rm PD} = 2.7 \pm 2.0\%$ at a ${\rm PA} = -29^{+22}_{-23} \degr$ in the FB and a  ${\rm PD} = 2.2 \pm 1.1\%$ at a ${\rm PA} = 37^{+15}_{-16} \degr$ in the SA, and a  ${\rm PD} = 0.7 \pm 0.6\%$ at a ${\rm PA} = 46^{+34}_{-26} \degr$ in the NB. The model-independent studies carried out for GX~349$+$2 show a similar variation of polarizations (considering the uncertainties) with a ${\rm PD} = 4.4 \pm 2.9\%$ at a ${\rm PA} = -33\pm18 \degr$ in the FB,  a  ${\rm PD} = 1.7 \pm 0.9\%$ at a ${\rm PA} = 39\pm15 \degr$ in the SA, and a ${\rm PD} = 2.4 \pm 1.6\%$ at a ${\rm PA} = 45\pm19 \degr$ in the NB.  Figure \ref{mod_ind_pol_br} shows that the energy-dependent branch-resolved polarimetry of GX~349$+$2 provides a slight hint of PA rotation of $\sim 60 \degr$ in the FB (albeit to be taken cautiously considering those are $<$ 2$\sigma$ estimations, see Table \ref{table2}) with respect to the SA/NB.

 Furthermore, our spectro-polarimetric studies of GX~349$+$2 show that the upper limits obtained on the polarization estimates of the individual components decrease slightly with our assumptions of the disk PA being either parallel or orthogonal to the PA of the Comptonized component. A further assumption of an unpolarized blackbody component attributed to the symmetric emissions from the NS surface results in a relatively smaller upper limit of the polarization of the individual components. With the current data, we can not favor one scenario over the other statistically. However, combining the spectral and polarimetric behavior of GX~349$+$2, if we consider an unpolarized blackbody component representing the direct emission from the NS surface, then, a Comptonized component with a slab-like corona, along with a relatively soft disk emission \citep{2022MNRAS.514.2561G}. However, we note here that in the case of  GX~349$+$2, although the previously reported spectral studies of the source along with the results reported in this paper favor a possible slab-like coronal geometry, we do not see any strong energy-dependent variations of PD, which could be inherent to Sco-like sources.

The polarization variations along the different states of the Z track have been reported for multiple Z sources (see Table~\ref{table8}). Our study shows a possible indication of similar PA rotation in the FB with respect to the SA/NB state. However, we note here that the results of higher PD detected in the FB relative to the SA/NB in the case of  GX~349$+$2 with an indication of PA rotation by 60\degr are not statistically significant, and more observations of GX~349$+$2 with simultaneous broadband coverage are necessary to investigate such variations of polarization along the Z track of GX~349$+$2.

\section{Summary}
\label{summary}
This work reports the first X-ray polarization study of the Z source NS LMXB GX~349$+$2. The X-ray spectro-polarimetric study using IXPE and NuSTAR shows a source polarization of  ${\rm PD} = 1.1 \pm 0.3\%$  with a polarization angle of ${\rm PA} = 32 \pm6 \degr$. The detailed spectro-polarimetric study shows the presence of a softer accretion disk component ($\sim$0.62-0.70~keV), assumed to be a source of the seed photons for Comptonization in a hot plasma with electron temperature $\text{kT}_{e} \sim$ 2.46-2.62 keV. The blackbody radiation component with temperature $\sim$ 1.29–1.30 keV represents possibly an unpolarized emission from the NS surface. The NuSTAR spectra show an asymmetric iron emission line ($\sim$ 6.7 keV) consistent with previous reports of the detections of reflection features from the source. Our studies demand the requirement of future multi-epoch IXPE observations with longer exposures, including simultaneous observations using X-ray instruments with better sensitivity and broad energy coverage. Our future radio observations would provide a better scope to investigate the nature of GX~349$+$2 as a Sco-like source.

\section{Acknowledgments}
\begin{acknowledgments}
This research used data provided by the Imaging X-ray Polarimetry Explorer (IXPE), NICER (Neutron star Interior Composition Explorer), Nuclear Spectroscopic Telescope Array (NuSTAR) and distributed with additional software tools by the High-Energy Astrophysics Science Archive Research Center (HEASARC), at NASA Goddard Space Flight Center (GSFC). U.K. and T.J.M. acknowledge support by NASA grant 80NSSC24K1747. M.N. is a Fonds de Recherche du Quebec – Nature et Technologies (FRQNT) postdoctoral fellow. We would like to thank Fabio La Monaca and Alessandro Di Marco for coordinating the NuSTAR observations simultaneously with our IXPE observations and for notifying us that they had done so to increase the value of the IXPE observations for the community, which significantly helped our branch-resolved spectro-polarimetric analysis.
\end{acknowledgments}


%

\vspace{5mm}
\facilities{IXPE, NICER, NuSTAR}


\software{ixpeobssim (Baldini et al. 2022), xspec
(Arnaud 1996), HEASoft (Blackburn 1995)
          }





\bibliography{sample631}{}
\bibliographystyle{aasjournal}



\end{document}